\documentclass[10pt]{article}

\usepackage{graphicx}

\usepackage{amsmath}
\usepackage{amssymb}

\usepackage{cite}

\usepackage{lineno}

\usepackage{microtype}
\DisableLigatures[f]{encoding = *, family = * }

\usepackage[labelfont=bf,labelsep=period,justification=raggedright]{caption}

\date{}

\begin{document}

\begin{flushleft}
{\Large \textbf{Preliminary quantification of freely exploring {\it
Atta insularis}} }
\\
A. Reyes, G. Rodr\'iguez and E. Altshuler$^{ \ast}$
\\
Group of Complex Systems and Statistical Physics, Physics Faculty,
University of Havana, Havana 10400, Cuba
\\
$^\ast$ E-mail: ealtshuler@fisica.uh.cu
\end{flushleft}

\vspace{0.5cm}
\vspace{0.5cm}



Ants are social insects that typically use pheromone traces to
self-organize long foraging lines without any centralized
organization \cite{Holldobler1990}. However, a very basic question
arises: how a single ant explores new territories without any chemical
or ``topographical" clues? To the authors knowledge, the question is
unanswered in the literature. In this paper, we offer a preliminary
quantification of the free exploration of the Cuban leaf-cutter ant
{\it Atta insularis} (known as {\it bibijagua} in Cuba)
\cite{Altshuler2005,Noda2006,Nicolis2013,Tejera2014,Reyes2016}.

\begin{figure}[h]
\begin{center}
 \includegraphics[width=10cm]{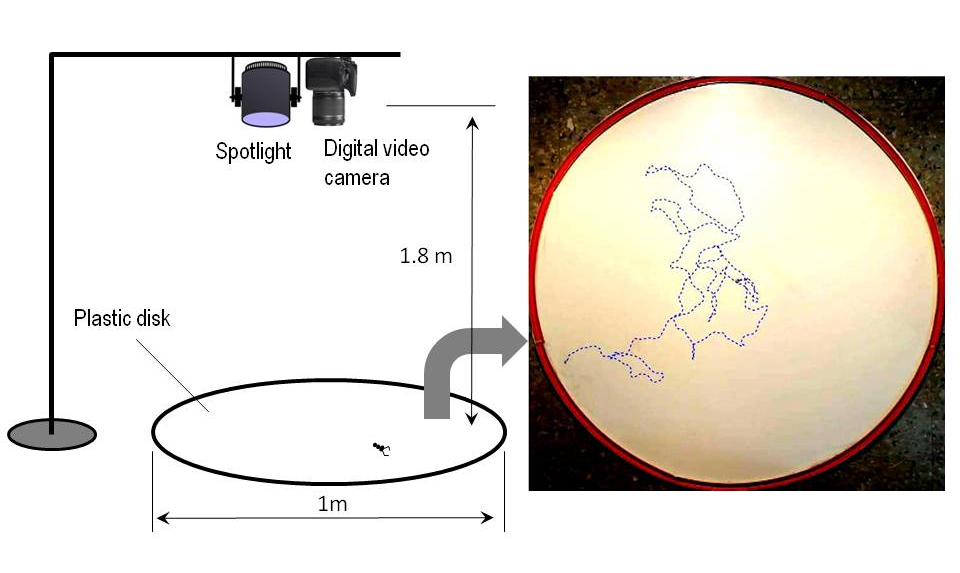}
 \caption{{\bf Experimental setup.} Left panel: sketch of the
experimental set up. The camera was a CANON EOS Rebel T3. Videos
were taken at 25 frames per second, with a resolution of 4272 pixels
$\times$ 2848 pixels. Right panel: Actual photogram of the
experimental arena, where an experimental trajectory of a single ant
is represented as a dotted line. The starting point of the
trajectory is at the geometrical center of the circular arena.}
 \label{Experimental schemme}
\end{center}
\end{figure}

A typical experiment can be described as follows. A single
individual of {\it Atta insularis} is collected from a foraging line
in natural conditions, immediately taken to the laboratory, and
released at the center of a 1-meter-diameter circular area made of
white plastic. The trajectory of the ant is then followed by a
digital camera located 1.8 meters above the center of the arena,
until the ant reaches the edge of the circular platform. The
illumination is provided by a 250-watt incandescent lamp located
near the camera, equipped with a light diffusor, as sketched in the
left panel of Fig.\ref{Experimental schemme}. Then, the individual
is released back in its natural foraging area. The experiment is
replicated for several individuals collected in the same way,
thoroughly cleaning the plastic arena with ethanol between
repetitions.

Images were averaged and the result was subtracted from each
photogram, in order to distinctly visualize the moving object (i.e.,
the ant), on an immobile and homogeneous background. Then, images were binarized
using an appropriately threshold, so the ant trajectory can be
easily converted into a list of positions and times. The right panel
of Fig.\ref{Experimental schemme} shows a typical trajectory.

\begin{figure}[h]
\begin{center}
 \vspace{0.5cm}
 \includegraphics[width=8.5cm]{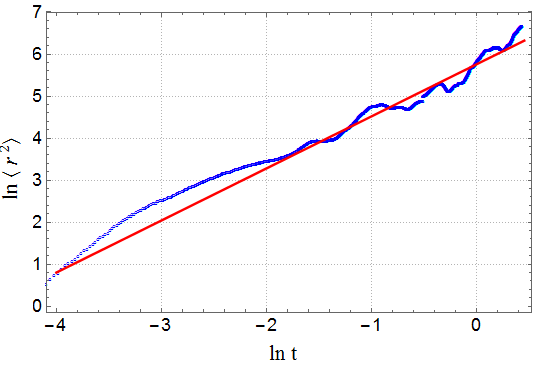}
 \caption{{\bf Mean Squared Displacement (MSD).} The figure shows the
logarithm of the MSD of the ants against the logarithm of time. The
graph comprises the data from a sample of 25 experiments with
individual ants. The straight line corresponds to a Log-Log graph of the
power law $\left\langle r{}^{2}\right\rangle\sim t^{\gamma}$ where the exponent
$\gamma=1.74 > 1$, suggests {\it super-diffusion}.}
\label{MSD}
\end{center}
\end{figure}

The first parameter we measured was the position of the ants as time
goes by, which was used to calculate the Mean Squared Displacement
(MSD), shown in Fig.\ref{MSD}. In the figure, it is included the data of 25
experiments (corresponding to 25 ants). The straight line fitting
the data can be described as $\left\langle r{}^{2}\right\rangle\sim t^{\gamma}$
with $\gamma=1.74$ (here, {\it r} is the absolute distance from the ants to the center of the arena).
Since the value of the exponent is bigger than 1, the motion
can be classified as {\it super-diffusive} --i.e., there is a
direction persistence beyond a pure random walk. Super-diffusive
behavior is expected when animals move in an ``anisotropic"
environment. For example, bacteria moving ``chemotactically" in a
chemical gradient of nutrients, or ``phototactically" in a light
gradient \cite{Berg2004}. In the case of ants, we would expect a
direction-biased walk if our arena was illuminated from one side
(ants would try to escape from light), but that is not the case. A
strongly super-diffusive behavior could be also expected if many
ants were released on the arena at the same time, since they would
chase each other due to the deposition of pheromone tracks --but
that is neither our case. The super-diffusive behavior we have found
in isolated ants might be related to ``direction memory" of the
individuals, or perhaps a short range (but not long range) lack of
memory (``Alzheimer walk", \cite{Viswanathan2011}).

\begin{figure}[h]
\begin{center}
 \vspace{0.5cm}
 \includegraphics[width=12cm]{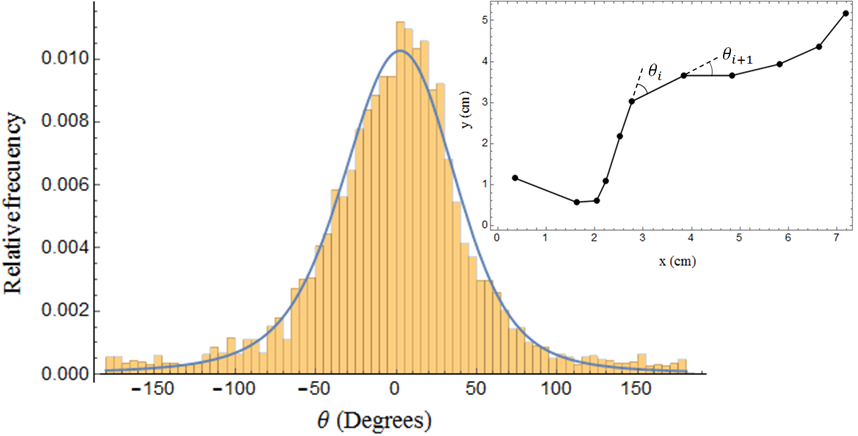}
 \caption{{\bf Frequency distribution of turn angles} The figure
represents the statistical distribution of turn angles, and comprise
the angles measured in all trajectories of all individual ants. The
inset indicates how the turn angle is defined.}
 \label{FrecuencyAngles}
\end{center}
\end{figure}

The second element we have computed is the statistics of turn
angles of the ant. The inset of Fig.\ref{FrecuencyAngles} illustrates how it was
measured. First, the trajectory of one ant was divided into straight
segments connecting two consecutive positions of the individual, separated
by a time interval of 0.8 seconds (20 frames). The turn angle was
defined as the deviation angle from one trajectory segment, to the
next.

The main frame of Fig.\ref{FrecuencyAngles} shows the statistical distribution of the
resulting angles, comprising all trajectories measured on 25
different experiments on 25 different ants. It can be immediately
seen the dominance of small turn angles: most of the time, ants
explore through smooth curves, and just eventually stop and perform
a major turn. A subtler observation is that the distribution is
slightly skewed to the left and the height of the right bars is considerable 
near zero, i.e., left turns are more probable (due to our
sign convention, it corresponds to a ``thicker" right side of the statistical
distribution shown in Fig.\ref{FrecuencyAngles}). That opens the question if ants are
``intrinsically left handed", so to speak.
Some authors have proposed that this asymmetry can result in
adaptive advantages. Studies with {\it Temnothorax albipennis}
ants show that they apparently use their right eye more the left eye to recognize
reference points \cite{Basari2014} (however, as we mentioned before, {\it Atta insularis} ants
are practically blind , so they are not likely to use any reference points
in our experiments). Another example of these antecedents
is illustrated in \cite{Hunt2014}, also with {\it Temnothorax albipennis}, showing more
left turns when ants explore new territories for the construction of a nest.

The continuous line in Fig.\ref{FrecuencyAngles} corresponds to a fit of a $\alpha$-stable
distribution (see, for example, \cite{Mandelbrot1960})which is able
to reproduce the ``power-law-like" tails of the bell-shaped
distribution.

During exploration, every few seconds ants stop and perform bigger
turns whose average is of the order of 90 degrees. In analogy with
the ``run-and-tumble" motion of {\it E. coli}
\cite{Berg2004,Altshuler2013,Figueroa2013,Figueroa2015} we define a
``tumble" as a bigger turn and a {\it run} as the fast
segment between two consecutive tumbles. Then, we have found that
the statistical distribution in the durations of runs follows an
exponential law, as ``classically" expected for bacteria. However,
more statistics must be collected to completely discard a power law.

In summary, we have preliminarily characterized the free exploration
of a social insect (the Cuban endemic ant {\it Atta insularis})
using parameters typically used in the area of micro-swimmers. On
the one hand, we have found a super-diffusive behavior in the
individual ants --which implies a certain level of direction
preference of unknown origin. Secondly, we have found clues
indicating that free-exploring ants are slightly ``left-handed",
i.e., they prefer to turn left instead of right during exploration.
These results --especially the latter-- must be corroborated by a
larger statistical sample of individuals, and should be extrapolated
to other, non-social, insect species.

\bibliographystyle{naturemag}


%
%
%

%

%
%
%

%
%


\end{document}